\def\g5{\gamma_5}

\documentstyle[preprint,prl,aps,epsfig]{revtex}  
\begin{document}  
\tightenlines
\title{Measurement of the Neutral Weak Form Factors of the Proton}  
  
\newcommand{\calstate}{1}  
\newcommand{\clermont}{2}  
\newcommand{\hampton}{3}  
\newcommand{\harvard}{4}  
\newcommand{\infnroma}{5}  
\newcommand{\tjnaf}{6}  
\newcommand{\kent}{7}  
\newcommand{\kentucky}{8}  
\newcommand{\kharkov}{9}  
\newcommand{\kyungpook}{10}  
\newcommand{\maryland}{11}  
\newcommand{\mitlns}{12}  
\newcommand{\newhamp}{13}  
\newcommand{\norfolk}{14}  
\newcommand{\nccu}{15}  
\newcommand{\olddom}{16}  
\newcommand{\princeton}{17}  
\newcommand{\regina}{18}  
\newcommand{\rutgers}{19}  
\newcommand{\saclay}{20}  
\newcommand{\stonybrook}{21}  
\newcommand{\syracuse}{22}  
\newcommand{\temple}{23}  
\newcommand{\wandm}{24}  
  
\author{K.~A.~Aniol,$^{\calstate}$  
D.~S.~Armstrong,$^{\wandm}$  
M.~Baylac,$^{\saclay}$  
E.~Burtin,$^{\saclay}$  
J.~Calarco,$^{\newhamp}$  
G.~D.~Cates,$^{\princeton}$  
C.~Cavata,$^{\saclay}$  
J.-P.~Chen,$^{\tjnaf}$  
E.~Chudakov,$^{\tjnaf}$  
D.~Dale,$^{\kentucky}$  
C.~W.~de~Jager,$^{\tjnaf}$  
A.~Deur,$^{\tjnaf,\clermont}$  
P.~Djawotho,$^{\wandm}$  
M.~B.~Epstein,$^{\calstate}$  
S.~Escoffier,$^{\saclay}$  
L.~Ewell,$^{\maryland}$  
N.~Falletto,$^{\saclay}$  
J.~M.~Finn,$^{\wandm}$  
K.~Fissum,$^{\mitlns}$  
A.~Fleck,$^{\regina}$  
B.~Frois,$^{\saclay}$  
J.~Gao,$^{\mitlns}$  
F.~Garibaldi,$^{\infnroma}$  
A.~Gasparian,$^{\hampton}$  
G.~M.~Gerstner,$^{\wandm}$  
R.~Gilman,$^{\rutgers}$  
A.~Glamazdin,$^{\kharkov}$  
J.~Gomez,$^{\tjnaf}$  
V.~Gorbenko,$^{\kharkov}$  
O.~Hansen,$^{\tjnaf}$  
F.~Hersman,$^{\newhamp}$  
R.~Holmes,$^{\syracuse}$  
M.~Holtrop,$^{\newhamp}$  
B.~Humensky,$^{\princeton}$  
S.~Incerti,$^{\temple}$  
J.~Jardillier,$^{\saclay}$  
M.~K.~Jones,$^{\wandm}$  
J.~Jorda,$^{\saclay}$  
C.~Jutier,$^{\olddom}$  
W.~Kahl,$^{\syracuse}$  
D.~H.~Kim,$^{\kyungpook}$  
M.~S.~Kim,$^{\kyungpook}$  
K.~Kramer,$^{\wandm}$  
K.~S.~Kumar,$^{\princeton}$  
M.~Kuss,$^{\tjnaf}$  
J.~LeRose,$^{\tjnaf}$  
M.~Leuschner,$^{\newhamp}$  
D.~Lhuillier,$^{\saclay}$  
N.~Liyanage,$^{\mitlns}$  
R.~Lourie,$^{\stonybrook}$  
R.~Madey,$^{\kent}$  
D.~J.~Margaziotis,$^{\calstate}$  
F.~Marie,$^{\saclay}$  
J.~Martino,$^{\saclay}$  
P.~Mastromarino,$^{\princeton}$  
K.~McCormick,$^{\olddom}$  
J.~McIntyre,$^{\rutgers}$  
Z.-E.~Meziani,$^{\temple}$  
R.~Michaels,$^{\tjnaf}$  
G.~W.~Miller,$^{\princeton}$  
D.~Neyret,$^{\saclay}$  
C.~Perdrisat,$^{\wandm}$  
G.~G.~Petratos,$^{\kent}$  
R.~Pomatsalyuk,$^{\kharkov}$  
J.~S.~Price,$^{\tjnaf}$  
D.~Prout,$^{\kent}$  
V.~Punjabi,$^{\norfolk}$  
T.~Pussieux,$^{\saclay}$  
G.~Qu\'em\'ener,$^{\wandm}$  
G.~Rutledge,$^{\wandm}$  
P.~M.~Rutt,$^{\tjnaf}$  
A.~Saha,$^{\tjnaf}$  
P.~A.~Souder,$^{\syracuse}$  
M.~Spradlin,$^{\princeton,\harvard}$  
R.~Suleiman,$^{\kent}$  
J.~Thompson,$^{\wandm}$  
L.~Todor,$^{\olddom}$  
P.~E.~Ulmer,$^{\olddom}$  
B.~Vlahovic,$^{\nccu}$  
K.~Wijesooriya,$^{\wandm}$  
R.~Wilson,$^{\harvard}$  
B.~Wojtsekhowski$^{\tjnaf}$ \\  
(HAPPEX Collaboration) \\ }  
  
\address{$^{\calstate}$ California State University - Los Angeles,   
Los Angeles, CA 90032,}  
\address{$^{\clermont}$ LPC, Universit\`e Blaise Pascal/IN2P3,  
F-63177 Aubi\`ere
, France }  
\address{$^{\hampton}$ Hampton University, Hampton, VA 23668}  
\address{$^{\harvard}$ Harvard University, Cambridge, MA 02138}  
\address{$^{\infnroma}$ Istituto Nazionale di  
Fisica Nucleare, Sezione Sanit\`a, 00161 Roma, Italy}  
\address{$^{\tjnaf}$ Thomas Jefferson National Accelerator   
Laboratory, Newport News, VA 23606}  
\address{$^{\kent}$ Kent State University, Kent, OH 44242}  
\address{$^{\kentucky}$ University of Kentucky, Lexington, KY 40506}  
\address{$^{\kharkov}$ Kharkov Institute of Physics and Technology, Kharkov  
310108, Ukraine}  
\address{$^{\kyungpook}$ Kyungpook National University, Taegu 702-701, Korea}  
\address{$^{\maryland}$ University of Maryland, College Park, MD 20742}  
\address{$^{\mitlns}$ Massachusetts Institute of Technology,   
Cambridge, MA 02139}  
\address{$^{\newhamp}$ University of New Hampshire, Durham, NH 03824}  
\address{$^{\norfolk}$ Norfolk State University, Norfolk, VA 23504}  
\address{$^{\nccu}$ North Carolina Central University, Durham, NC 27707}  
\address{$^{\olddom}$ Old Dominion University, Norfolk, VA 23508}  
\address{$^{\princeton}$ Princeton University, Princeton, NJ 08544}  
\address{$^{\regina}$ University of Regina, Regina, SK S4S 0A2, Canada}  
\address{$^{\rutgers}$ Rutgers, The State University of New Jersey,   
Piscataway, NJ 08855}  
\address{$^{\saclay}$ CEA Saclay, DAPNIA/SPhN, F-91191   
Gif-sur-Yvette
, France }  
\address{$^{\stonybrook}$ State University of New York at   
Stony Brook, Stony Brook, NY 11794}  
\address{$^{\syracuse}$ Syracuse University, Syracuse, NY 13244}  
\address{$^{\temple}$ Temple University, Philadelphia, PA 19122}  
\address{$^{\wandm}$ College of William and Mary, Williamsburg, VA 23187}  
  
\date{\today}  
  
\maketitle  
  
\begin{abstract}  
We have measured the parity-violating electroweak asymmetry  
in the elastic scattering of polarized electrons from the  
proton.   The kinematic point ($\langle\theta_{\mathrm lab}\rangle=  
12.3^{\circ}$ and   
$\langle Q^2\rangle=0.48 $ (GeV/c)$^2$) is  
chosen to provide sensitivity, at a level that is of theoretical interest,  
to the strange electric form factor $G_E^s$. The result,   
$A=-14.5\pm2.2$ ppm,  
is consistent with the electroweak Standard Model and no additional  
contributions from strange quarks. In particular, the measurement implies  
$G_E^s+0.39G_M^s = 0.023\pm 0.034$ (stat) $\pm \; 0.022 $ (syst)   
$\pm \; 0.026\ (\delta G_E^n)$, where the last uncertainty arises from the  
estimated uncertainty in the neutron electric form factor.  
\end{abstract}  
\draft 
 
The proton, which is believed to be a state of three quarks bound by the  
strong  
force of QCD, is a complex object when probed at intermediate  
energies.  In order to develop a useful description, one must first  
establish all the relevant degrees of freedom. Recent theoretical and  
experimental investigations have indicated that strangeness might play an  
important role\cite{KM,REV}.   
For example, do $s\overline s$ pairs contribute to the charge radius  
or magnetic moment of the proton?  Quite possibly, since the mass  
of the strange quark is comparable to the proton mass and  
the scale of the strong interaction.  On the other hand, the empirically  
successful OZI rule predicts that the effects of strange quarks are  
greatly suppressed at low energies\cite{GI}.  Resolution of this issue   
requires that it be addressed experimentally.  
  
A particularly clean experimental technique~\cite{SMB}   
for isolating the effects  
of strange quarks in the nucleon is measuring parity-violation amplitudes  
in the elastic scattering of polarized electrons   
from protons~\cite{BM}.  The theoretical asymmetry, which is   
caused by the interference between the weak and electromagnetic amplitudes,   
is given in the Standard Model by~\cite{REV}  
  
\begin{equation}  
{{A}_{th}}=\frac{\sigma_R-\sigma_L}{\sigma_R+\sigma_L}=\Biggl[  
\frac{-G_FQ^2}{\pi\alpha\sqrt{2}}\Biggr]  
\label{eq:apv}  
\end{equation}  
\[  
\times\frac{\varepsilon G^{p\gamma}_EG_E^{pZ}+\tau G^{p\gamma}_MG_M^{pZ}-  
\frac{1}{2}(1-4\sin^2\theta_W)\varepsilon'G_M^{p\gamma}G_A^{pZ}}  
{\varepsilon(G^{p\gamma}_E)^2+\tau(G^{p\gamma}_M)^2}  
\]  
where $G_{E}^{p\gamma}(G_{M}^{p\gamma})$ is the electric(magnetic)   
Sachs form factor for  
photon exchange, $G_{E,M}^{pZ}$ is the corresponding  
quantity for Z$^0$ exchange and $\theta_W$ is the electroweak mixing  
angle.  All form factors are functions of $Q^2$ and $\varepsilon,\ \tau,$ and  
$\varepsilon'$ are kinematic quantities (see Ref. \cite{SAM}).    
For our kinematics, $\tau\sim0.136,$ $\varepsilon\sim.97$,  
$\varepsilon'\ll 1$   
and the term involving $G_A^{pZ}$ contributes only a few percent relative to  
the other terms.    
The predicted asymmetry is on the order of 10 parts per million (ppm).  
  
To interpret the experiment, $G^{p,Z}_{E,M}$ can be expressed in terms  
of proton, neutron, and strange form factors if  
the up(down) quarks in the proton have the same properties as  
the down(up) quarks in the neutron (assumption of isospin  
symmetry).  Then  
\[
G^{p,Z}_{E,M}=\frac{1}{4}( G^{p\gamma}_{E,M} -G^{n\gamma}_{E,M})  
-\sin^2\theta_W G^{p\gamma}_{E,M}   
-\frac{1}{4}G^s_{E,M}   
\]
and, if the electromagnetic form factors are sufficiently well known from   
experiment, the only unknown quantities involve strange form factors.  
  
Extensive literature is devoted to estimating  
the size of strange form factors.  
Approaches include~\cite{REV,GI,RLJ,MB,WEI,HAM,MI} pole fits,  
meson loops, the NJL model, vector dominance, unquenched quark model,  
chiral symmetry, and skyrme models.    
The significance of the strange form factors  
is attested to by the fact that they are relevant to many  
theoretical approaches striving to understand QCD at low energies.  
Some of the calculations predict substantial effects, 50\% or more of the   
asymmetry at our kinematics, that are dominated by $G_E^s$.   
Other calculations predict effects at the few percent level or less.    
The goal of our  
experiment is to determine if indeed the strange quark form factors  
are large enough to be an important part of any detailed  
description of the proton.  
  
The experiment took place in Hall~A  
at the Thomas Jefferson National Accelerator  
Facility (JLab).  A $\sim 100\mu$A continuous-wave beam of   
longitudinally polarized  
3.356 GeV electrons was scattered from  
a 15 cm long liquid hydrogen target.    
The electrons which were scattered elastically at   
$\langle\theta_{\mathrm lab}\rangle\sim \pm 12.3^{\circ}$   
were focussed by two identical high resolution   
5.5 msr spectrometers onto a total-absorption detector made up of a  
lead-lucite  
sandwich.  Only the scattered electrons were detected; the second  
spectrometer  
merely doubled the solid angle.    
The spectrometers, which deflect the electrons by 45$^{\circ}$ out of the  
scattering plane, focus inelastic trajectories   
well away from our detectors.  
  
The polarized electron beam originated from a bulk GaAs photocathode excited  
by circularly polarized laser light.    
The helicity of the beam was set every 33.3 ms locked to the 60 Hz  
frequency of the AC power in the lab.  The helicity was structured  
as pairs of consecutive 33.3 ms periods with opposite helicity,   
henceforth called  
windows.  The helicity of the first window in each pair was determined by a   
pseudo-random number generator.  All signals were integrated  
over a 32 ms gate which began $\sim 1$ ms after the start of each window.  
The output of  the integrators was digitized by 16-bit customized analog   
to digital converters.  Integration and digitization were handled  
by custom-built modules designed to minimize noise and crosstalk.  
  
The experimental method is driven primarily by the fact that the measured  
asymmetry is a few ppm. The trick to  
measuring small asymmetries is to maintain negligible  
correlations between the helicity of the beam and any other  
properties of the beam, such as intensity, energy, position, or  
angle.  At JLab, the only quantity for which we found a non-zero  
helicity-correlated difference was intensity.  
This correlation was reduced to below 1 ppm  
by using a slow feedback system.  
  
The intensity of the beam was measured with two independent RF  
cavities and the position of the beam was measured at five  
locations with RF stripline monitors.    
Window-to-window jitter in the intensity was typically 300 ppm and  
window-to-window jitter in position was a few microns.  This  
impressive stability of the accelerated beam made it easy to set stringent  
limits on any helicity-correlated beam parameters.  Averaged over the  
entire run, limits on the position differences were typically on the order  
of   
a few nm.  
One of the position monitors was located at a point of high dispersion  
in the transport line and set a limit on the average helicity-correlated  
fractional energy difference at the $10^{-8}$ level.  
  
Limits on the impact of the helicity correlations  
were determined by modulating the beam position and energy  
concurrent with data taking.  Since these changes were small and  
uncorrelated with helicity, the same data could be used both for  
calibration purposes as well as for the primary data sample.  
The results of these studies show that the contribution of the correlations  
to the raw asymmetry is a factor of 20 smaller than the statistical   
error and thus negligible.    
  
\begin{figure}[t!]  
\vspace{-1cm}  
\epsffile{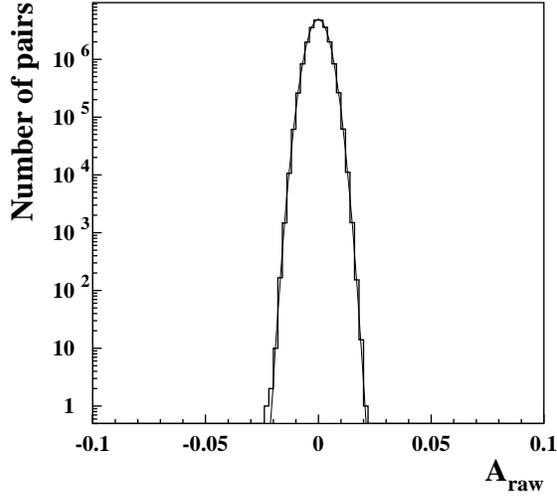}  
\caption[Pairs]{Distribution of $A_{raw}$ for individual window  
pairs.  Only data with $I>80\mu$A ($\sim 95\%$ of sample) are shown.    
The curve is a Gaussian fit.}  
\label{fig:allp}  
\end{figure}

The raw asymmetry for each window pair is defined   
\[A_{raw}=[(D_R/I_R)-(D_L/I_L)]/  
[(D_R/I_R)+(D_L/I_L)],\]  
where $D_R(D_L)$ is the detector signal and $I_R(I_L)$  
is the signal from the intensity cavity for the right(left) helicity window.  
Pedestals for these signals were measured during beam-off periods and the  
linearity of the system was  
verified at the 1\% level during periods when the beam current was ramping  
up.  
A histogram of the distribution of window pair asymmetries  
is given in Fig. \ref{fig:allp}.  The distribution is purely Gaussian.  
Separate auxiliary tests (at lower beam energy and thus high cross  
section) carried out prior to the run demonstrated  
that boiling of the liquid target did not increase the noise in the  
asymmetry measurements.  
The only cuts applied to our data sample were when the beam was $< 3\mu$A  
or when equipment such as the spectrometer or target was clearly  
malfunctioning. 
 
\begin{figure}[t!]  
\epsffile{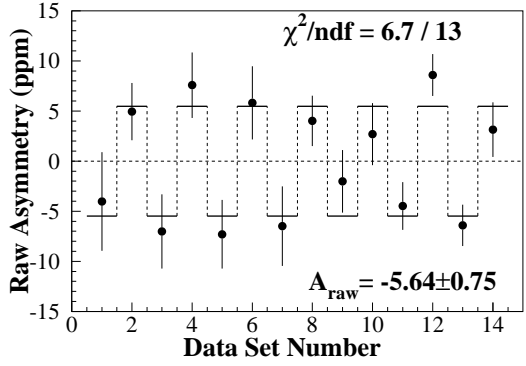}  
\caption[Slug]{Average values of $A_{raw}$ for each data set.  
Odd data sets have the $\lambda/2$ plate inserted in the laser beam.  
The plate is expected simply to change the sign of the raw asymmetry.}  
\label{fig:slug}  
\end{figure}

One important test for the presence of false asymmetries is through the  
insertion of a half-wave ($\lambda/2$) plate in the laser beam.    
This complements  
the helicity of the electron beam and hence the sign of the raw  
asymmetry while leaving many other possible systematic effects unchanged.  
The data were taken in sets of 24-48 hr duration, and the $\lambda/2$  
plate was inserted for the odd numbered sets.  The raw asymmetries for  
each set are given in Fig. \ref{fig:slug}.    
A clear correlation between the presence of the   
$\lambda/2$ plate and the sign of the asymmetry is seen. Since   
the target is unpolarized, this correlation is an unambiguous signal  
of parity violation.  
Averages of the raw asymmetries for the entire data set, representing  
78~C of electrons on target, are given in Table I for both $\lambda/2$  
plate settings and for each spectrometer individually.  The  
results for the subsets are consistent with each other.  
  
To extract the experimental asymmetry $A_{exp}=A_{raw}/P_e$,   
the beam polarization was measured both by Mott scattering  
near the injector and by M{\o}ller scattering just upstream of the  
hydrogen target.  We use the average value $P_e=(0.388\pm0.027).$  
The $Q^2$ of the data, averaged over the acceptance of the detector,  
was determined to be $0.479\pm0.003$ (GeV/c)$^2$  
by separate low-current runs that used  
tracking drift chambers in front of our detectors to study individual  
events.  The drift chambers were used also to measure possible   
inelastic backgrounds from pole-tip scattering  
by varying the central momentum of the spectrometers so that the  
dominant elastic events would follow the trajectories of inelastic  
events under running conditions.  The contribution of background  
to our asymmetry is at most 2\% of 15 ppm as listed in the summary of  
errors in Table II.       
The result is $A_{exp}=-14.5\pm2.0  
(stat)\pm1.1(syst)$ ppm.

To study the effects of strange quarks, we compare our  
result with $A_{th}$ (Eq. \ref{eq:apv}) using parameterizations of  
the form factors.  
For $G_E^n$ we use the function due to Galster~\cite{GAL}.   
The difference between the true value and the Galster approximation is  
denoted $\delta G_E^n$.  
It is estimated to be $\pm$50\% of the Galster function,  
corresponding to a 9.6\% error in $A_{th}$.  
We will leave this as a separate error since it is significant and since  
experiments in progress should improve the value of $\delta G_E^n$.  
For the other form factors, the dipole parameterization is  
taken as a reasonable approximation at our $Q^2$:  $G^p_E=G_D$,  
$G^p_M=\mu_pG_D$, and $G^n_M=\mu_nG_D$\cite{REV}.  
This introduces an uncertainty in the predicted asymmetry of   
about 4\% of itself.  Electroweak 
radiative corrections~\cite{REV,PDG}, which are known  
and only on the order of a few percent of the asymmetry, were applied.  
The kinematic  
suppression of the $G_A^Z$ term is essential in our experiment to control the  
otherwise large radiative corrections in that term.    
With these assumptions, $A_{th}=-15.8\pm0.7\pm1.5(\delta G_E^n)$ ppm.  
  
\begin{figure}[htbp]  
\epsffile{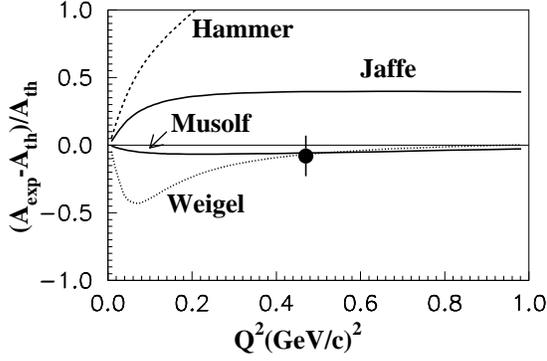}  
\caption[Results.]{Experimental $\delta A/A$  
assuming $\delta G_E^n=0$, together  
with representative theoretical calculations by Jaffe~\cite{RLJ},  
Hammer {\it et al.}~\cite{HAM}, Musolf and Burkhardt~\cite{MB}, and  
Weigel~\cite{WEI}.  For papers that  
did not include the $Q^2$ dependence \cite{RLJ,MB} a dipole form  
is assumed as suggested in Ref. \cite{REV}.}  
\label{fig:result}  
\end{figure}

\begin{figure}[htbp]  
\epsffile{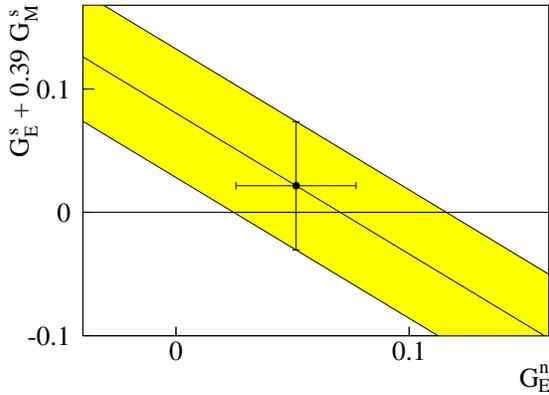}  
\caption[G.]{Allowed region of space in $ (G^s_E+0.39G^s_M)$  
versus $ G_E^n$ at $Q^2=0.48$ (GeV/c)$^2$.  Data point assumes  
Galster approximation for $G_E^n$.}  
\label{fig:svsn}  
\end{figure}

Representative calculations for $\delta A=(A_{exp}-A_{th})/A_{th}$  
are given in Fig. \ref{fig:result} together  
with our data point under the assumption that $\delta G_E^n$ is negligible.  
The largest of the predictions are excluded by our data.  Previous  
data sensitive to different combinations of the form factors  
and at different $Q^2$ values are also consistent with the absence  
of strange quarks, but at a somewhat less sensitive level\cite{SAM,GAR}.  
From our data, we can extract the combination of strange form factors  
at $Q^2=0.48$ (GeV/c)$^2$:   
$G_E^s+0.39G_M^s = 0.023\pm 0.034$ (stat) $\pm \; 0.022 $ (syst)   
$\pm \; 0.026\ (\delta G_E^n)$.  
Our result is shown in Fig. \ref{fig:svsn},   
expressed as the combination of strange  
form factors that we measure versus $G_E^n$.

We plan to improve our precision by a factor of 2 in 1999.  Improvements  
in $G_E^n$ will be important for us to extract useful information.  
Although we have ruled out some of the more generous predictions, it  
is important to pursue the subject further.  Expanding the $Q^2$ range  
is important, as well as separating $G_E^s$ from $G_M^s$, either by    
varying the kinematics or by using an isoscalar target such as $^4$He.  
  
The relative ease with which we were able to measure the small   
asymmetry at JLab bodes well for the future of experiments  
measuring parity-violating amplitudes.  The high quality of the  
beam provided by this new facility is invaluable for the  
performance of precision experiments.  
  
We wish to thank the entire staff at JLab for their tireless  
work in developing this new facility, and particularly  
C. K. Sinclair and M. Poelker for their timely work on the  
polarized source.  This work was supported, in part, by the Department of  
Energy, the National Science Foundation, the Korean Science  
and Engineering Foundation (Korea), the INFN (Italy), the Natural 
Sciences and Engineering Research Council of Canada, and the  
Commissariat \`a l'\'Energie Atomique (France).

\begin{table}  
  
\begin{tabular}{lrrr}  
&\multicolumn{1}{c}{$\lambda/2$ out} &  
\multicolumn{1}{c}{$\lambda/2$ in} & Combined\\ \hline  
det1	&	$5.1\pm1.4$	&	$-3.3\pm1.6$	&$-4.25\pm1.06$\\  
det2	&	$6.2\pm1.5$	&	$-8.1\pm1.6$&$-7.07\pm1.07$\\ \hline  
Total	&	$5.6\pm1.0$	&	$-5.6\pm1.1$	&$-5.64\pm0.75$\\   
\end{tabular}  
\caption{Averages of $A_{raw}$ (in ppm).  The different spectrometers 
are det1 and det2.}  
\end{table}  
  
\begin{table}  
  
\begin{tabular}{llr}  
$A$&Source of error & \multicolumn{1}{c}{$\Delta A/A$(\%)}\\	\hline  
{$A_{raw}\ \ \ $}&	Statistics		&	13.4\\   
		&	Others			&	$<$0.3\\ \hline  
{$A_{exp}$}	&	Beam Polarization	&	7\\  
		&	$Q^2$ Determination	&	1\\  
		&	Backgrounds		&	2\\ \hline  
{$A_{th}$}	&	Nucleon Form Factors&\\  
&\multicolumn{1}{r}{(excluding $G_E^n$)}	&	4.0\\  
		&	Radiative Corrections	&	1.4\\  
		&	$G_E^n$			&	9.6\\	\hline  
\end{tabular}  
\caption{Summary of contributions to the errors for $A_{raw},\ A_{exp}$, and  
$A_{th}$.}  
\end{table}

\epsfxsize=8cm  
  
\epsfysize=6cm  

\epsfxsize=8cm  
\epsfysize=6cm  

\epsfysize=6cm  

\end{document}